\documentclass[
    ,final            
  ]
  {aipproc}

\layoutstyle{8x11single}

\begin{document}

\title{Prototyping an Active Neutron Veto for SuperCDMS}

\classification{07.05.Fb;29.30.Ep;29.40.Mc}
\keywords      {shielding, SuperCDMS, active neutron veto}

\author{Robert Calkins}{
  address={Department of Physics, Southern Methodist University, Dallas, Texas 75275, U.S.A. }
}

\author{Ben Loer}{
  address={Fermi National Accelerator Laboratory,Batavia, Illinois 60510, U.S.A. }
}

\begin{abstract}
Neutrons, originating cosmogenically or from radioactive decays, can produce signals in dark matter detectors that are indistinguishable from Weakly Interacting Massive Particles (WIMPs). To combat this background for the SuperCDMS SNOLAB experiment, we are investigating designs for an active neutron veto within the constrained space of the compact SuperCDMS passive shielding. The current design employs an organic liquid scintillator mixed with an agent to enhance thermal neutron captures, with the scintillation light collected using wavelength-shifting fibers and read out by silicon photo-multipliers. We will describe the proposed veto and its predicted efficiency in detail and give some recent results from our R\&D and prototyping efforts.
\end{abstract}

\maketitle


\section{Introduction, SuperCDMS SNOLAB and Basic Neutron Detection Technique}
The SuperCDMS SNOLAB experiment will consist of eight towers of six crystal detectors each, cooled down to ~10s of mK temperatures. Each detector will be manufactured from either high purity germanium or silicon and instrumented 
with phonon and ionization sensors on each side. 
The initial payload will consist of six towers of six Ge detectors each, one tower consisting of six Si detectors and one tower of four Ge and two Si detectors biased at high voltage.
The detectors will be housed within an ultra pure copper cryostat which provides thermal 
protection as well as some shielding from external gamma rays. The external shielding will consist of layers of lead, high density polyethylene and water tanks for the initial deployment at SNOLAB. The polyethylene layer will be
removed and replaced with an active neutron veto during a possible future upgrade phase of the experiment.

The ionization yield for a given phonon energy is a powerful discriminating variable between electron recoils and nuclear recoils in an iZIP detector\cite{Agnese:2013rvf}. 
Nuclear recoils have a lower yield with respect to electron recoils. We expect a WIMP to interact with the target nucleus and produce a nuclear recoil signature. 
Neutrons can also interact with the target nucleus, producing a nuclear recoil which can be a problematic source of background for many WIMP searches. 
We expect the neutron background to be produced from trace amounts of U and Th contamination in the detector construction materials from spontaneous fission and alpha-n production.
We also expect a from a very small neutron contribution originating from cosmogenic production. 
The ability to veto on these events would reduce the neutron background in WIMP search experiments.

\begin{figure}[h!]
  \includegraphics[height=.2\textheight]{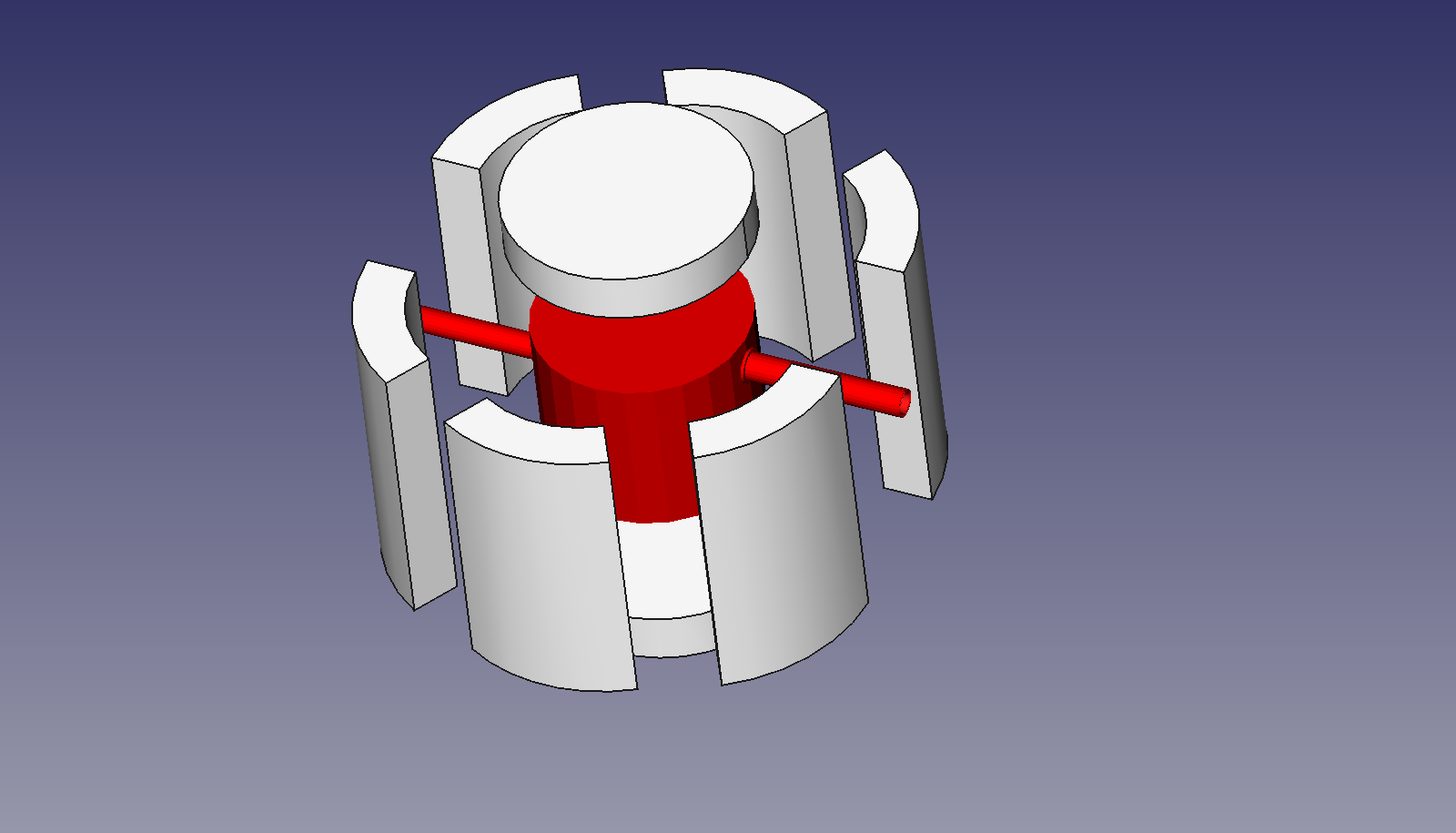}
  \caption{Exploded view of the proposed active neutron veto surrounding the SuperCDMS SNOLAB cryostat. Each module will be instrumented with the exception of the base module located underneath the cryostat. }
	\label{SNOLAB}
\end{figure}

$^{10}$B has a large thermal neutron capture cross-section and is often added to polyethylene to enhance the neutron capture efficiency. The $^{10}$B-neutron capture reaction produces a $^7$Li atom along with an alpha. The majority
of the interactions produce the $^7$Li in an excited state which releases a gamma as it decays down to the ground state. These decay products produce light in the liquid scintillator that will be detectable but 
unfortunately the output is heavily quenched so we expect to see energies around 50-60 ke$V_{ee}$\cite{Wright:2010bu}.

The veto design consists of acrylic tank modules surrounding the cryostat as shown in Fig. \ref{SNOLAB}. Each module will be constructed with acrylic and filled with a liquid scintillator mixture consisting of 
70\% linear alkylbenzene(LAB) which is an organic scintillator and 30\% trimethylborate (TMB) that acts as a neutron capture target. 
The organic scintillator mixture will be doped with 2,5-Diphenyloxazole (PPO) at a 2 g/L concentration and 1,4-Bis(2-methylstyryl)benzene (bisMSB) at 6.5 mg/L.
Light will be collected from each module through wavelength shifting fibers (WLS) strung through the module and delivered to silicon photo-multipliers (SiPMs). SiPMs have some advantages over PMTs for our application since 
they generally have lower intrinsic radioactivity and have a compact size. 
The base module will be uninstrumented high density polyethylene for structural support.

\section{Liquid Scintillator Prototype}
\begin{figure}[h!]
  \includegraphics[height=.2\textheight]{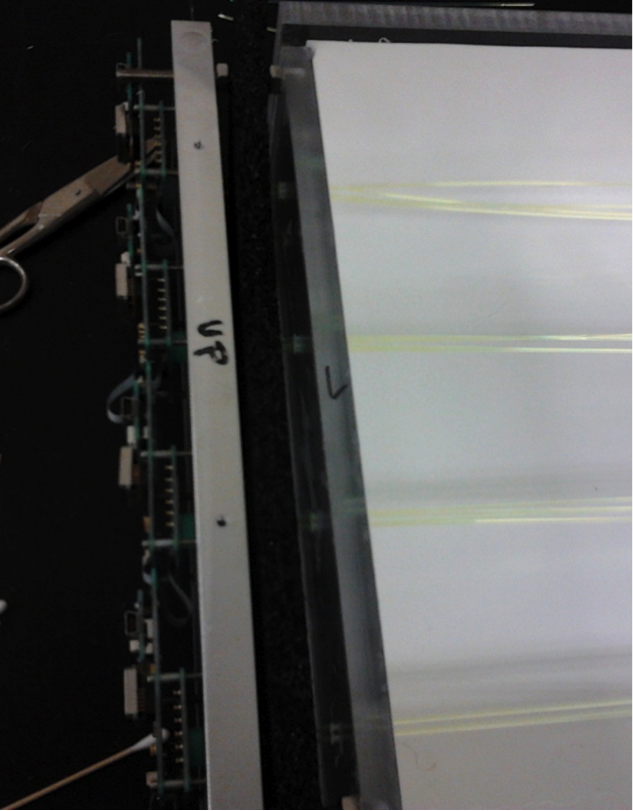}
  \includegraphics[height=.2\textheight]{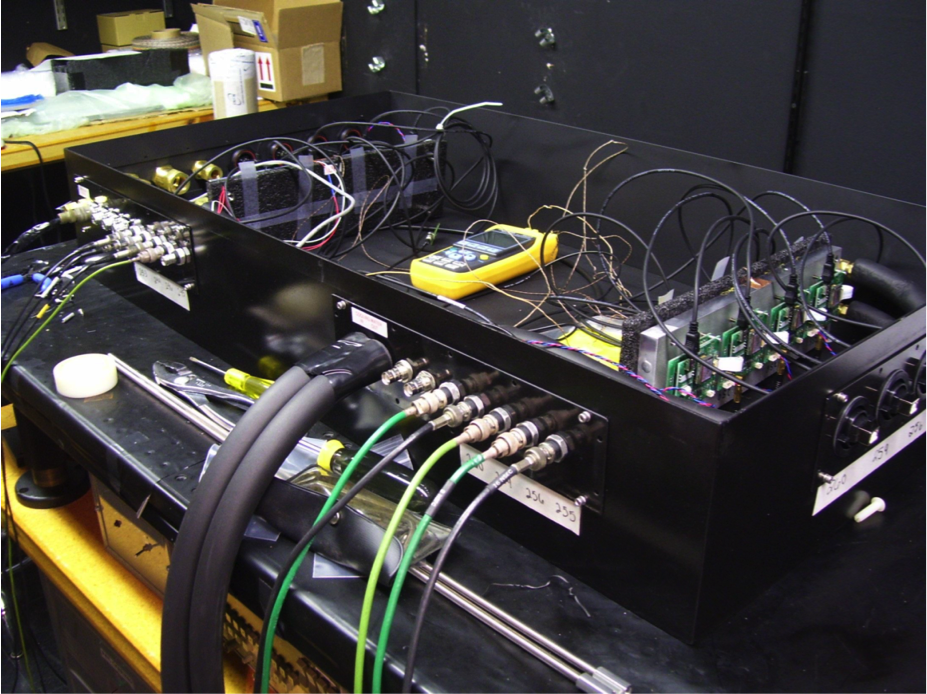}
  \caption{(Left) This photograph shows one end of the liquid scintillator prototype before it was sealed shut. The SiPMs are mounted to the Al cooling plate and are linked to the fibers that are immersed in the liquid scintillator. 
(Right) The measurements were conducted in dark box that was held under nitrogen purge. This picture shows the dark box after the SiPMs were installed but before the acrylic vessel was inserted. }
	\label{box}
\end{figure}

The prototype dimensions were chosen to correspond to approximately one quarter the size of a single full size module. A box was used instead of the curved module geometry to simplify construction and assembly. The prototype
dimensions were 23 x 12 x 3.75 inches and was assembled out of 0.5 inch thick clear acrylic. 
To increase the light collection efficiency, the interior of the acrylic vessel were lined with Lumirror reflector, as can be seen in Fig. \ref{box}.

The prototype was strung with 16 1.5 mm diameter Kuraray Y-11 WLS fibers through holes drilled through the acrylic and sealed with 5-minute epoxy. Each fiber was 28 inches long and were bundled into groups of four at each end. 
The fiber readout was accomplished using eight Hamamatsu S12572-100C 3 mm SiPMs, one for each end of the 4 4-fiber bundles. Custom mounts were constructed to mount the fiber ends over the face of the SiPM. The SiPMs were mounted 
on a cooling plate in order to reduce the dark rate.

The entire assembly was housed in a dark box, as shown in Fig. \ref{box} right, under a fume hood for safety. The atmosphere in the dark box was held under a nitrogen purge to reduce the flammability hazard of the TMB and to prevent condensation
from forming on the SiPMs.

\section{Results with the Liquid Scintillator Prototype}

Data was collected with the liquid scintillator prototype between June and October 2014 for a variety of sources. 
The sources were used to calibrate the SiPM response and the light yield of the prototype. 
Based on the measured light yields, the energy deposition from a neutron capture should be above the noise threshold for the prototype. We exposed the prototype to a $^{252}$Cf source but were unable to observe any neutron captures.
The
reason is unknown but the large gamma background originating from the source maybe overwhelming the capture signal. It may also be that the energy resolution was too poor to definitively resolve the neutron capture peak.

\begin{figure}[h!]
  \includegraphics[height=.3\textheight]{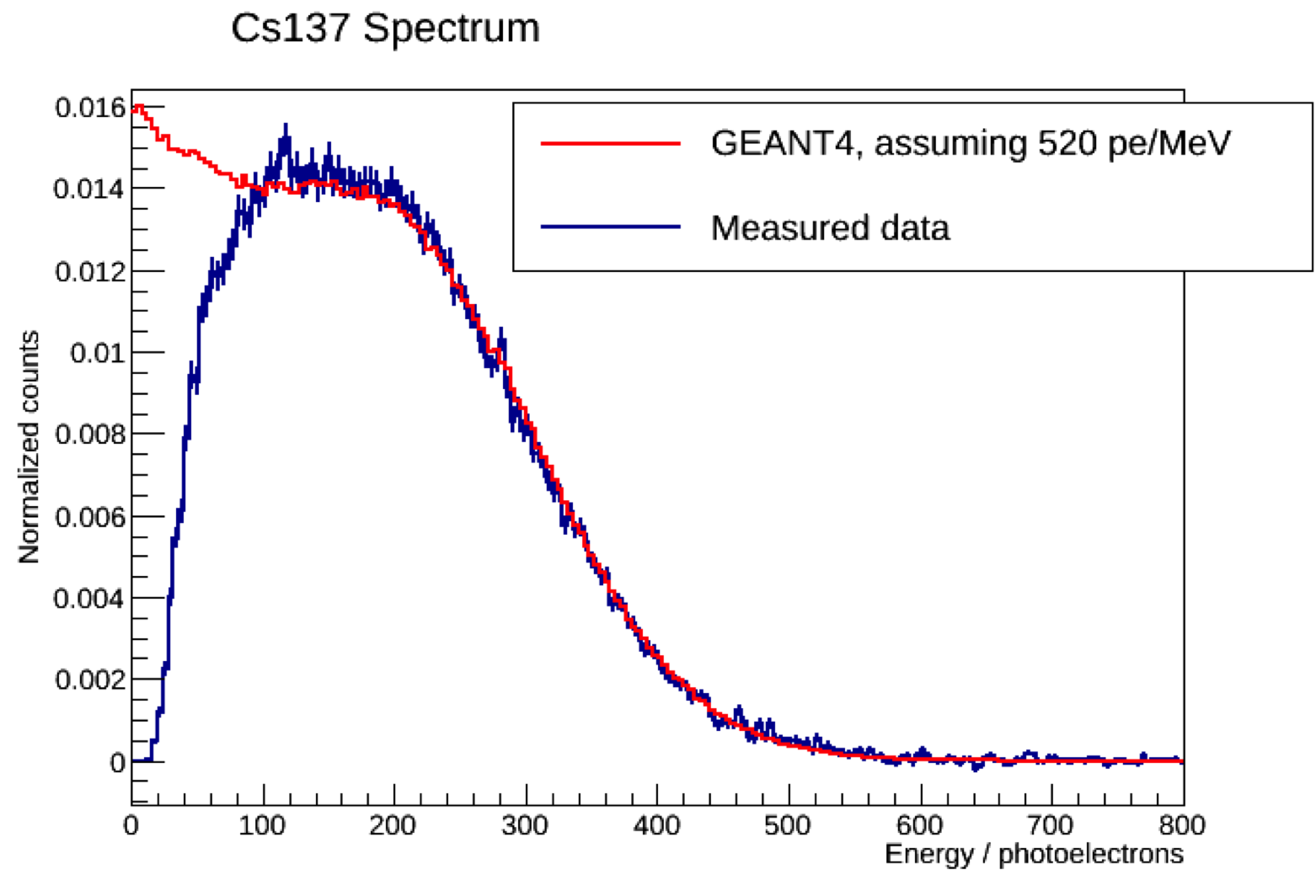}
  \caption{This figure shows the measured Cs137 spectrum. The Geant4 spectrum was smeared using parameters derived from the Cs137 data and SiPM response\cite{Agostinelli:2002hh}. 
}
	\label{cs137}
\end{figure}

\begin{figure}[h!]
  \includegraphics[height=.3\textheight]{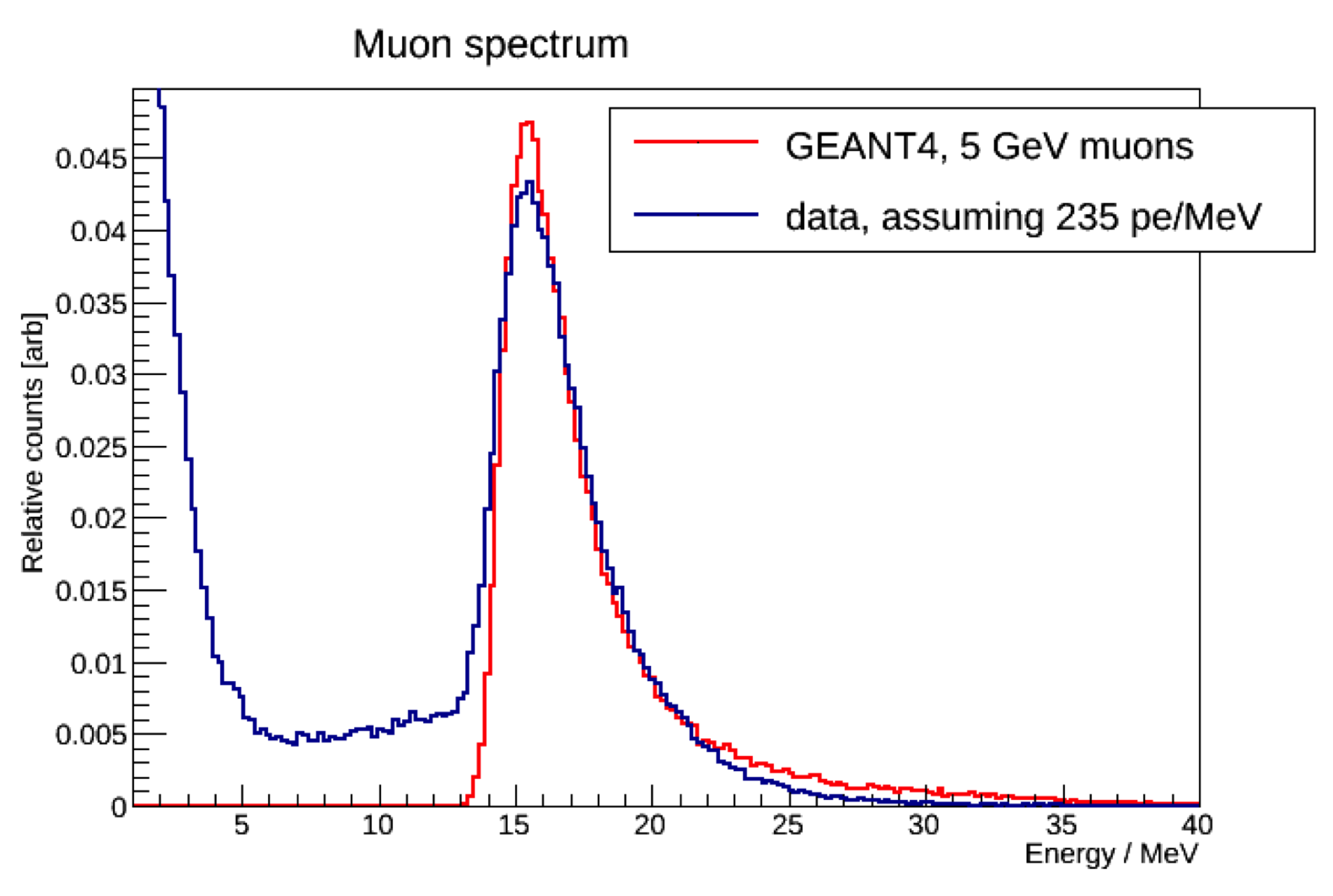}
  \caption{This figure shows the measured energy spectrum for data collected without any radioactive source present. 
The peak around 16 MeV corresponds to cosmic muons passing through the detector. 
The Geant4 energy spectrum
has been convoluted with a smearing function derived from Cs137 data. 
The energy scale for the data has been set by aligning the data and Monte Carlo peaks by eye.
}
	\label{muon}
\end{figure}

\section{Plastic Scintillator Option}
\begin{figure}[h!]
  \includegraphics[height=.3\textheight]{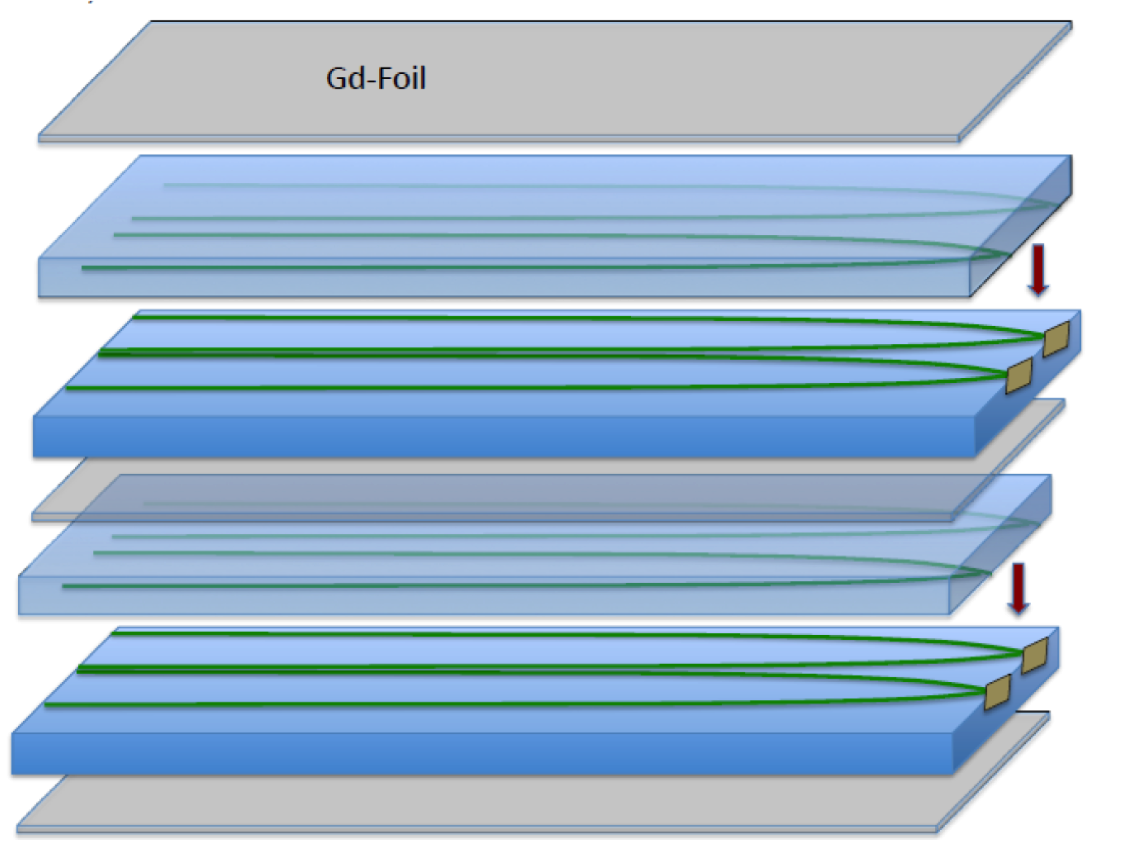}
  \includegraphics[height=.3\textheight]{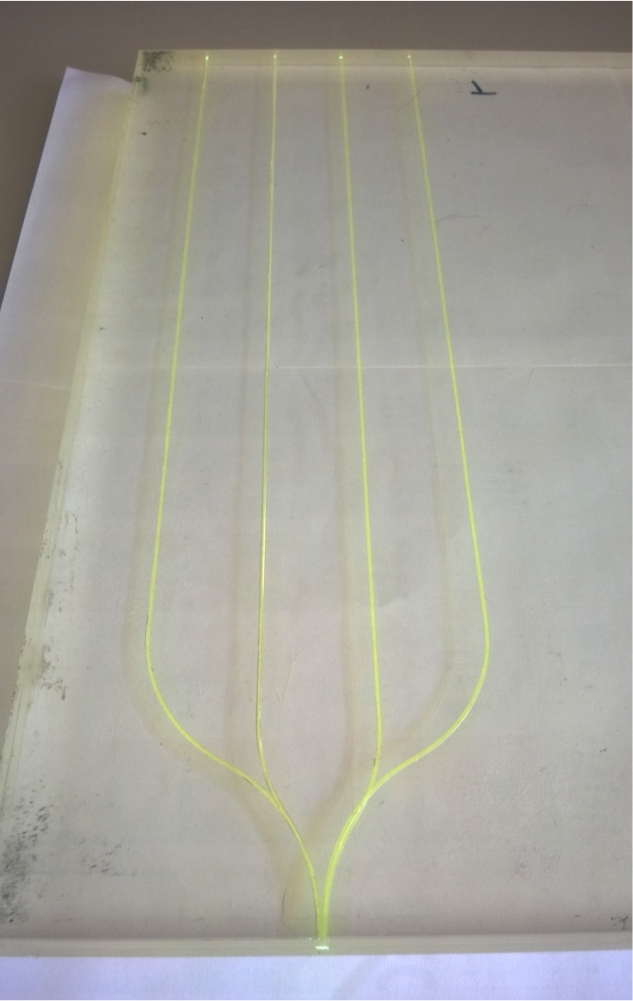}
  \caption{(Left)A potential alternative to inserting boron or gadolinium into a plastic scintillator is using thin sheets instead. The neutron capture happens in the doped sheets and the products are measured in a typical
plastic scintillator layer. (Right) This picture shows one prototype scintillator panel. The WLS fiber has been embedded between two panels that have had groves routed to accommodate the fiber. }
\label{plastic}
\end{figure}

Plastic scintillators are an alternative to the liquid scintillator design that is also being considered. The liquid scintillator design relies on TMB, which is flammable and poses a fire hazard. It requires a significant
investment in fire suppression infrastructure at SNOLAB which will place a burden on the experiment's budget. Since it is a liquid, there are also hazards associated with spilling and filling. In terms of infrastructure, solid
plastic modules can be easier to work with and avoid many of the problems of the LAB+TMB mixture. A large drawback is that it is difficult to put TMB into a plastic scintillator while simultaneously maintaining good optical
properties. One potential solution to this problem would be to use Gd doped thin foils or paints to provide a thermal neutron capture target and interleave this with plastic scintillator modules as illustrated in Fig. \ref{plastic}.

\section{Summary}
The success of the SuperCDMS SNOLAB WIMP search depends on sufficiently controlling backgrounds. We have presented a design for an active neutron veto as a possible upgrade for the SuperCDMS SNOLAB experiment. The baseline design
utilizes acrylic tanks of liquid scintillator doped with boron to enhance the neutron capture cross-section. 
Results from
a one quarter scale prototype have been presented but fail to show a definitive neutron capture peak for unknown reasons. An alternative design, that avoids some of the issues of the liquid scintillator design by 
utilizing plastic scintillators, was also presented.


\begin{theacknowledgments}
The authors would like to thank Anna Pla-Dalmau, Abaz Kryemadhi, Katrina Schrock and Mathew Bressler. 
This material is based upon work supported by the DOE, Universities Research Association and National Science Foundation under Grant Number (1151869). Any opinions, findings, and conclusions or recommendations expressed in this material are those of the author(s) and do not necessarily reflect the views of the National Science Foundation.
Fermilab is operated by the Fermi Research Alliance, LLC under Contract No. DE-AC02-07CH11359 with the United States Department of Energy.
\end{theacknowledgments}



\bibliographystyle{aipproc}   

\bibliography{LRT}

\IfFileExists{\jobname.bbl}{}
 {\typeout{}
  \typeout{******************************************}
  \typeout{** Please run "bibtex \jobname" to optain}
  \typeout{** the bibliography and then re-run LaTeX}
  \typeout{** twice to fix the references!}
  \typeout{******************************************}
  \typeout{}
 }

\end{document}